\documentclass[10pt,conference]{IEEEtran}
\IEEEoverridecommandlockouts
\usepackage{cite}
\usepackage{booktabs}
\usepackage{caption}
\usepackage{amsmath,amssymb,amsfonts}
\usepackage{algorithmic}
\usepackage{graphicx}
\usepackage{textcomp}
\usepackage{xcolor}
\usepackage{subcaption}
\usepackage{float}
\usepackage{hyperref}
\usepackage{multirow}

\def\BibTeX{{\rm B\kern-.05em{\sc i\kern-.025em b}\kern-.08em
    T\kern-.1667em\lower.7ex\hbox{E}\kern-.125emX}}
\begin{document}

\title{AI for Requirements Engineering: Industry Adoption and Practitioner Perspectives\\

\author{
\IEEEauthorblockN{
Lekshmi Murali Rani, 
Richard Berntsson Svensson, 
Robert Feldt
}
\IEEEauthorblockA{
\textit{Department of Computer Science and Engineering} \\
\textit{Chalmers University of Technology and University of Gothenburg, Sweden} \\
%SE-41296, Gothenburg , Sweden\\
lekshmi@chalmers.se, richard@cse.gu.se, robert.feldt@chalmers.se
}
}

}

\maketitle

\begin{abstract}
The integration of AI for Requirements Engineering (RE) presents significant benefits but also poses real challenges. Although RE is fundamental to software engineering, limited research has examined AI adoption in RE. We surveyed 55 software practitioners to map AI usage across four RE phases: Elicitation, Analysis, Specification, and Validation, and four approaches for decision-making:human-only decisions, AI validation, Human–AI Collaboration (HAIC), and full AI automation. Participants also shared their perceptions, challenges, and opportunities when applying AI for RE tasks. Our data show that 58.2\% of respondents already use AI in RE, and 69.1\% view its impact as positive or very positive. HAIC dominates practice, accounting for 54.4\% of all RE techniques, while full AI automation remains minimal at 5.4\%. Passive AI validation (4.4–6.2\%) lags even further behind, indicating that practitioners value AI’s active support over passive oversight. These findings suggest that AI is most effective when positioned as a collaborative partner rather than a replacement for human expertise. It also highlights the need for RE-specific HAIC frameworks along with robust and responsible AI governance as AI adoption in RE grows.

\end{abstract}


\section{Introduction}

Requirements engineering (RE) is regarded as the foundation upon which successful projects are created and includes four key phases, namely elicitation, analysis, specification, and validation~\cite{b17}. The direct impact of requirements quality and completeness on project outcomes makes RE a critical phase for AI integration consideration~\cite{b16}. The emergence of large language models (LLMs) and AI-powered tools has created new opportunities for RE tasks due to their ability for natural language processing, extraction, and analysis of large volumes of data, and automation of repetitive tasks~\cite{b18}. 

Recent research has explored AI applications in RE with a focus on algorithmic performance and technical feasibility rather than real-world adoption patterns~\cite{b2,b19}. While there are multiple studies on AI-assisted RE, e.g.,~\cite{b20,b21}, there exists a significant gap in understanding how practitioners are integrating AI tools into their daily RE processes, what collaboration patterns are used between humans and AI models, and what challenges and opportunities emerge during these integrations in real-world contexts. In addition, it is also important to consider responsible AI usage in RE tasks as RE involves high-stakes decision-making that can directly impact software functionality, user experience, and business outcomes. 

This study aims to explore the current state and extent of AI adoption in the field of RE, how AI tools/models (e.g., LLMs) are currently being used or could be used as collaborators or assistants in RE phases, perceptions, challenges, risks, and barriers to integrating AI in various RE tasks and the future potential for AI-based enhancements across RE processes such as elicitation, specification, analysis, and validation. The study categorizes AI usage into four approaches based on interaction patterns and decision authority: humans make decisions (human-only decisions), AI validates human decisions (AI validation), AI assists/collaborates with humans (Human–AI Collaboration), and AI handles processes individually (full AI automation). Human–AI Collaboration (HAIC) specifically refers to bidirectional interactions where both human expertise and AI capabilities contribute to RE outcomes. This study addresses three research questions as follows: the current state of AI adoption in RE practices and the extent of AI integration across RE phases (RQ1), practitioners' perception of using AI for RE (RQ2), and the challenges and opportunities of using AI for RE (RQ3). 

To address these questions, a survey was conducted with 55 practitioners across diverse roles, industries, and organizational contexts. The survey captured AI usage patterns among practitioners, providing statistical generalizability while also offering a detailed exploration of the challenges and opportunities as perceived by the practitioners. The survey examined AI usage patterns across the four core phases of RE, responsible AI practices among adopters, and insights into the challenges and opportunities practitioners face when integrating AI into their requirements work. The main contribution of this study is a comprehensive empirical overview of AI adoption patterns in RE phases, providing insights into practitioners' perceptions, challenges, and opportunities for using AI for RE.

The remaining section of this paper is organized as follows. Section 2 presents Related Work. Section 3 describes the Research Methodology for this study. Section 4 presents the Results, while Section 5 discusses the Results. Section 6 concludes the paper.

\section{Related Work}

Studies show that LLM can be used effectively in various RE phases, including elicitation, specification, analysis, and validation (e.g.,~\cite{b1,b22}). Recent research shows that while GPT series models dominate current applications by 67.3\% of studies, existing architectures face technical challenges including interpretability (61.9\%), reproducibility (52.4\%), and controllability (47.6\%)~\cite{b2}. Studies suggest using LLM to extract semantic relationships between abstract representations of system capabilities, to improve automated reasoning techniques for eliciting and analyzing the consistency of normative requirements (e.g,~\cite{b10}). Requirements elicitation frequently uses AI techniques, mainly NLP and ML, which extract requirements from digital data sources such as forums and micro-blogs and textual data, such as user feedback~\cite{b6,b7}. Advanced LLMs (e.g., GPT-4 and Claude 3.5 Sonnet) were used to verify requirements against system specifications, achieving high accuracy in identifying non-fulfilled requirements, indicating their potential for requirements verification tasks~\cite{b9}. ChatGPT-generated requirements were found to be highly abstract, atomic, consistent, correct, and understandable compared to human-formulated requirements~\cite{b4}. LLMs can assist in generating UML sequence diagrams from natural language requirements, although challenges remain in ensuring completeness and correctness~\cite{b5}.

While these studies show promising algorithmic capabilities in requirements generation, verification, and modeling, they mainly focus on proof-of-concept implementations rather than addressing the fundamental technical challenges of interpretability, reproducibility, and controllability that limit practical deployment~\cite{b2}. Also, issues related to ambiguity and inconsistency in requirements remain unresolved even in high-performing models~\cite{b4}. LLMs can help create interaction models and textual use cases for complex systems, but to increase their precision and reliability, more refinement and domain-specific training are required~\cite{b8}. Integrating LLMs into existing RE workflows presents challenges, including the need for domain-specific knowledge, handling ambiguity, and ensuring data quality~\cite{b6}. Current tool development efforts lack systematic approaches for integrating AI capabilities into established RE workflows. Systematic reviews show that governance-related issues (e.g., ethics) form a distinct cluster of challenges that need solutions, but they are addressed by less than 20\% of studies in AI applications~\cite{b2}. The literature shows that AI governance guidelines are difficult to operationalize due to context-specific knowledge gaps, insufficient professional training in ethics, and minimal integration between engineering-focused and ethics-focused research.

Current literature mainly examines algorithmic capabilities rather than understanding how practitioners currently use AI for RE processes. There is no systematic way of transferring AI technologies into the practice of RE, highlighting the need for a better understanding of practitioner needs and current usage patterns. While HAIC is recognized as important, systematic approaches to understanding how practitioners envision AI as a collaborative partner in RE processes also remain underexplored.

\section{Research Methodology}
The main objective of this study was to investigate the current use, perceptions, practices, and challenges with the integration of AI tools and models in the RE process in the industry. This study examined how AI is currently used in various RE phases, namely elicitation, specification, analysis, and validation, and what factors influenced its adoption or rejection in organizational settings. This study investigated the following three research questions:

\noindent\textbf{RQ1:} \textit{What is the current extent and nature of AI adoption in RE practices, and how does the AI integration pattern vary across different RE phases?}\\
\textbf{RQ2:} \textit{What are the perceptions of practitioners regarding using AI for RE?}\\
\textbf{RQ3:} \textit{What are the challenges and opportunities in using AI for RE?}

\noindent\textbf{Survey Design and Validation:} To answer the research questions, this study employed a quantitative research design using a survey research approach, with an online survey comprising 39 questions. A descriptive-exploratory survey design was used to investigate the emerging area of using AI for RE to identify current practices and perceptions. The survey included a mix of multiple-choice, Likert-scale, and open-ended questions on (a) the demographics (b) current use of AI tools/models in RE and overall perception about AI usage or adoption (c) RE activity-specific integration (elicitation, analysis, specification and validation) (d) risks, challenges and adoption barriers (e) responsible AI practices for RE. The survey instrument was developed in accordance with areas of interest and related literature, and involved pilot testing with two participants. The full questionnaire, data analysis code, quantitative data analysis details, and the thematic analysis codebook are available in \href{https://doi.org/10.5281/zenodo.15648803}{Zenodo}~\cite{b23}.

\noindent\textbf{Data Collection:} We administered the survey using the \href{https://www.soscisurvey.de/}{So-sci} survey platform to ensure secure data collection and participant anonymity. It was available for a period of 84 days in 2025. Informed consent was obtained from all participants before beginning the survey. The target group for the survey was software practitioners, including software developers, project managers, business analysts, and other practitioners in the software engineering (SE) life cycle. The participants included AI users (practitioners using AI tools for SE tasks) and non-AI users (practitioners not using them). The survey used a convenience sampling approach where it was distributed across multiple channels, including professional Slack groups, LinkedIn groups, and to personal contacts working in the software industry. The multi-channel distribution strategy was used to reach a diverse population of software practitioners across different roles, industries, and geographical locations.

\noindent\textbf{Quantitative Data Analysis:} Quantitative data analysis involved descriptive statistics for closed-ended questions. Before the analysis process, the data were checked for missing values and outliers, and missing data were handled by excluding them from the analysis. Descriptive statistics were used to analyze participant demographics (role distribution, industry distribution, years of experience, team size, development methodology), and AI adoption patterns (current AI usage and AI perception). The responsible AI factors and RE technique-specific AI adoption were also analyzed using descriptive analysis methods. A comparative analysis was also done to identify the HAIC patterns across RE phases, phase-specific challenges and opportunities, and to understand the consistency of practitioner preferences across different RE phases. 

\noindent\textbf{Qualitative Data Analysis:}
Open-ended responses were analyzed using inductive thematic analysis using Braun and Clarke’s approach~\cite{b24}, whereby themes were generated from the data itself rather than being guided by pre-existing theories or frameworks. It provided insights into the challenges and opportunities of AI usage in RE phases, offered specific examples of AI integrations, and highlighted barriers to AI adoption. The coding decisions are documented in the thematic analysis codebook with quotes and themes~\cite{b23}. 

\vspace{5pt}
\noindent\textbf{Threats to Validity:} As participants voluntarily chose to participate, there might be an over-representation of people who are either very interested in AI or have strong opinions on it, causing self-selection bias (Internal validity threat). To mitigate this, multi-channel distribution was used to ensure more diverse participants and positioned the survey as a study on "current use" of AI, targeting both users and non-users of AI. Self-reported data about AI usage may be subject to response bias (Internal validity threat). We also did not differentiate between types of AI tools used, which may vary significantly in capabilities and application (Internal validity threat). Our convenience sample (N = 55) may overrepresent AI enthusiasts and may not fully represent the broader RE practitioners (limiting generalizability), while the cross-sectional data cannot track adoption over time (External validity threat). To mitigate this, we ensured role coverage, industry diversity, experience, and team size variation. In addition, the rapid evolution of AI tools may lead to the findings becoming quickly outdated, leading to temporal validity limitations (External validity threat). To mitigate this, the main focus was on the RE process rather than on specific AI tools, where the findings can remain relevant even when the tools evolve. In addition, all four main RE phases were covered to create a baseline for longitudinal comparisons. Different organizations may define RE phases differently, which could affect the responses (Construct validity threat). To mitigate this, clear definitions were provided for each phase, and technique lists were provided based on industry terminology, along with flexible response options. Finally, the cross-sectional nature of our study captures only a snapshot of rapidly evolving AI adoption patterns (Construct validity threat).

\section{Results}
This section presents the results of the survey organized according to the three research questions in Section III.

\subsection{\textbf{Participant Demographics and Context}}
The survey collected responses from 55 SE practitioners from diverse roles and industries. Software developers constituted the largest group (30.9\%), followed by project leads/managers (21.8\%), as shown in Table~\ref{role}. The participants showed substantial professional experience with a mean of 9.6 years (median: 8.0 years, range: 1-35 years) and worked in teams averaging 19.9 members (median: 10, range: 2-150). The automotive industry was most represented (16.4\%), followed by finance (12.7\%) and retail (9.1\%), as shown in Table~\ref{industry}. Agile methodology dominated the development approaches (81.8\%), with hybrid approaches accounting for 12.7\% and plan-driven methods for 5.5\%. 

\begin{table}[ht]
\centering
\footnotesize
\caption{Participant Demographics}
\begin{subtable}{0.48\columnwidth}
\centering
\caption{Role Distribution}
\label{role}
\begin{tabular}{lr}
\toprule
\textbf{Role} & \textbf{Count (\%)} \\
\midrule
Software Developer & 17 (30.9\%) \\
Project Lead / Manager & 12 (21.8\%) \\
Business Analyst & 3 (5.5\%) \\
Product Manager & 3 (5.5\%) \\
Data Engineer & 3 (5.5\%) \\
Software Architect & 2 (3.6\%) \\
Support Engineer & 2 (3.6\%) \\
Engineering Manager & 2 (3.6\%) \\
Requirements Engineer & 1 (1.8\%) \\
AI Engineer & 1 (1.8\%) \\
\bottomrule
\end{tabular}
\end{subtable}
\hfill
\begin{subtable}{0.48\columnwidth}
\centering
\caption{Industry Distribution}
\label{industry}
\begin{tabular}{lr}
\toprule
\textbf{Industry} & \textbf{Count (\%)} \\
\midrule
Automotive & 9 (16.4\%) \\
Finance & 7 (12.7\%) \\
Retail & 5 (9.1\%) \\
Education & 4 (7.3\%) \\
Healthcare & 3 (5.5\%) \\
Manufacturing & 3 (5.5\%) \\
Technology & 3 (5.5\%) \\
E-commerce & 2 (3.6\%) \\
Security & 2 (3.6\%) \\
Public Sector & 2 (3.6\%) \\
\bottomrule
\end{tabular}
\end{subtable}

\label{tab:demographics}
\end{table}

\subsection{\textbf{RQ1: Current state of adoption and integration of AI in RE practices}}

\noindent\textbf{Adoption Levels:} Current AI adoption in RE shows a significant usage pattern, with 58.2\% (N = 32) of respondents reporting active use of AI tools in the RE processes. Among AI users, frequency of usage varies by phase: elicitation (59\% occasional, 22\% frequent, 16\% most of the time, 3\% never use), analysis (41\% occasional, 25\% frequent, 16\% most of the time, 19\% never use), specification (47\% occasional, 31\% frequent, 16\% most of the time, 3\% never use), and validation (47\% occasional, 34\% frequent, 9\% most of the time, 9\% never use).

\noindent\textbf{AI integration levels and Phase specific patterns:} HAIC dominates across all RE phases, representing 49.2\% to 60.5\% of all the techniques used, as shown in Table~\ref{phase_approach}. Analysis shows the highest AI collaboration (60.5\%) and lowest human-only decision-making (31.3\%), suggesting this phase benefits most from AI's analytical capabilities. Elicitation maintains higher human decision-making (39.2\%), showing the inherently communicative and interpretive nature of stakeholder interaction that requires human judgment and emotional intelligence. The specification phase has the highest autonomous AI use (7.6\%), possibly because documentation and diagram generation can be more easily automated. The validation phase reveals balanced patterns similar to Elicitation, underscoring the critical role of quality assurance, which necessitates human judgment. Full AI automation remains consistently low (3.8\%-7.6\%).

\begin{table*}[ht]
\centering
\caption{Decision-Making approaches across four key RE Phases}
\label{phase_approach}
\begin{tabular*}{\textwidth}{|l@{\extracolsep{\fill}}|c|c|c|c|}
\hline
\textbf{Decision-Making Approach} & \textbf{Elicitation} & \textbf{Analysis} & \textbf{Specification} & \textbf{Validation} \\
\hline
Humans make decisions & 39.2\% of phase activity & 31.3\% of phase activity & 32.8\% of phase activity & 34.9\% of phase activity \\
\hline
AI validates human decisions & 6.1\% of phase activity & 4.4\% of phase activity & 5.5\% of phase activity & 6.2\% of phase activity \\
\hline
AI assists/collaborates with humans & 49.2\% of phase activity & 60.5\% of phase activity & 54.1\% of phase activity & 53.6\% of phase activity \\
\hline
AI handles process individually & 5.4\% of phase activity & 3.8\% of phase activity & 7.6\% of phase activity & 5.3\% of phase activity \\
\hline
\hline
\textbf{Number of phase activities} & \textbf{12} & \textbf{11} & \textbf{14} & \textbf{15} \\
\hline
\end{tabular*}
\end{table*}

This shows that practitioners have largely moved beyond viewing AI as either a replacement for humans or just a validation tool, instead considering it as a collaborative partner, with the Analysis phase showing the highest collaboration rate (60.5\%). Human decision-making remains significant (31.3\%-39.2\%), while autonomous AI usage remains limited (3.8\%-7.6\%). These results suggest that organizations are adopting a collaborative approach to AI integration in RE phases, using the AI capabilities while maintaining human oversight for critical decision-making cases. The limited adoption of AI automation in RE phases shows caution regarding full AI autonomy in the RE process. This could be due to regulatory or compliance requirements, a lack of trust in current AI capabilities, or an understanding of RE complexity requiring human oversight.

\noindent\textbf{Responsible AI Practices Among Adopters:} Among 32 AI-adopting practitioners, responsible AI practices showed varied adoption levels:  Having humans review and approve AI suggestions was most prevalent (81.2\%), followed by allowing humans to correct or override AI-generated requirements (71.9\%) and checking AI-generated requirements for accuracy and reliability(68.8\%). Mid-level adoption was observed for training team members on how to use AI tools responsibly (59.4\%), ensuring AI follows data privacy and GDPR rules (59.4\%), and ensuring AI-generated recommendations are explainable and transparent (56.2\%). Lower adoption rates were found for assessing AI risk level (37.5\%), keeping records of AI-generated requirements (37.5\%), informing users when AI is assisting with requirements (37.5\%), and protecting AI tools and data from security risks (34.4\%). This pattern indicates reactive rather than proactive governance, with practitioners prioritizing immediate oversight over systematic risk management and auditability.

\subsection{\textbf{RQ2: Perceptions of practitioners on using AI for RE}}
\begin{figure}
    \centering
    \includegraphics[width=1\linewidth]{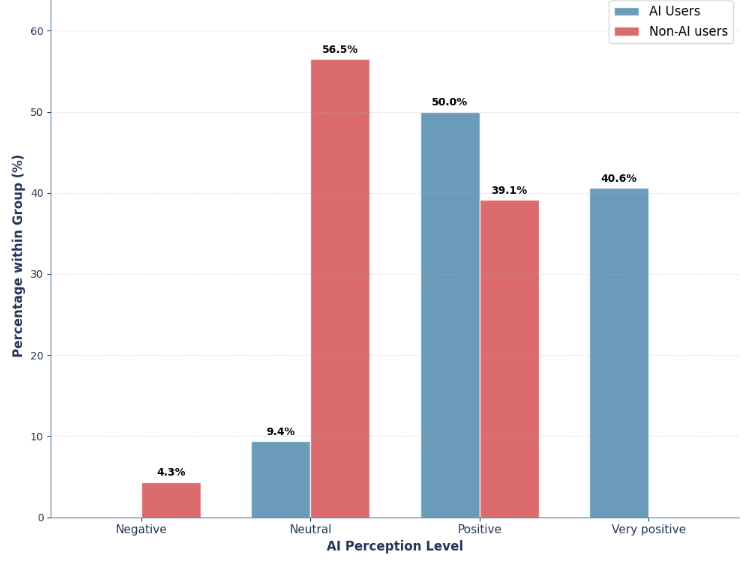}
    \caption{AI for RE Perception: User vs Non-AI Users}
    \label{perception}
    \vspace{-0.4cm}
\end{figure}

\noindent\textbf{Overall Attitudes:} Practitioners' attitudes toward AI in RE were highly positive: 45.5\% positive, 23.6\% very positive, 29.1\% neutral, with only 1.8\% negative perceptions. AI users (N=32) mostly have positive and very-positive feelings about AI, while non-AI users (N=23) mostly have neutral or positive feelings, as shown in Figure~\ref{perception}. This implies that AI usage can transform perception, as negative attitudes were found exclusively among non-AI users (4.3\%), while very positive attitudes were found only among AI users (40.6\%), indicating that experience using AI for RE can convert skeptical non-AI users and promote broader AI adoption.

\noindent\textbf{Phase-Specific Perceptions:} Table~\ref{ai_usage_perception} shows the AI usage perception across the four RE phases. Requirements elicitation shows the highest potential for automation with human oversight, with 36\% of the respondents feeling that AI can automate significant parts of the process but still requires human oversight, which is nearly triple the attitude seen in other phases (11-15\%). On the other hand, validation showed more caution with AI usage, with 32\% believing that AI can provide insights but cannot directly assist in decision-making. Analysis (20\%) and specification (22\%) showed approximately twice the collaboration potential by actively assisting in the process compared to elicitation (11\%) and validation (13\%). Full AI autonomy was mostly rejected, with only 2\% believing that AI could handle elicitation independently without human intervention, while no one believed that AI could handle analysis, specification, and validation independently without human intervention. AI's potential was recognized in the elicitation and validation phase, where no participants believed that AI contributes any value to the process, while uncertainty remains consistently high across all phases (26-29\% unsure/neutral), indicating that the RE field is still exploring practical AI applications. 

\noindent\textbf{Experience Effects:} AI users show four times higher confidence in automation with human oversight compared to non-AI users for all phases except validation. Non-AI users remain consistently uncertain about AI capabilities across all phases, with uncertainty levels 2-3 times higher than AI users. AI users show 2-4 times higher confidence in AI as collaborative partners, with the highest gaps in validation and analysis.

\begin{table*}[ht]
\renewcommand{\arraystretch}{1.3}
\caption{AI usage perception across all respondents vs AI users vs non-AI users across four phases}
\label{ai_usage_perception}
\centering
\footnotesize
\begin{tabular}{|p{5cm}|c|c|c||c|c|c||c|c|c||c|c|c|}
\hline
\multirow{3}{3.2cm}{\textbf{AI Usage Category}} & \multicolumn{3}{c||}{\textbf{Elicitation}} & \multicolumn{3}{c||}{\textbf{Analysis}} & \multicolumn{3}{c||}{\textbf{Specification}} & \multicolumn{3}{c|}{\textbf{Validation}} \\
\cline{2-13}
& \textbf{All} & \textbf{AI} & \textbf{Non-} & \textbf{All} & \textbf{AI} & \textbf{Non-} & \textbf{All} & \textbf{AI} & \textbf{Non-} & \textbf{All} & \textbf{AI} & \textbf{Non-} \\
& \textbf{} & \textbf{user} & \textbf{AI} & \textbf{} & \textbf{user} & \textbf{AI} & \textbf{} & \textbf{user} & \textbf{AI} & \textbf{} & \textbf{user} & \textbf{AI} \\
\hline
\hline
Can automate significant parts but requires human oversight & 36\% & 53\% & 13\% & 11\% & 16\% & 4\% & 13\% & 19\% & 4\% & 15\% & 16\% & 13\% \\
\hline
Unsure/Neutral about AI capability & 27\% & 16\% & 43\% & 29\% & 16\% & 43\% & 28\% & 16\% & 39\% & 26\% & 19\% & 30\% \\
\hline
Can collaborate with humans by actively assisting in the process & 11\% & 13\% & 9\% & 20\% & 29\% & 9\% & 22\% & 29\% & 13\% & 13\% & 19\% & 4\% \\
\hline
Tool for validation and enhancement of decision-making & 11\% & 9\% & 13\% & 15\% & 16\% & 13\% & 13\% & 13\% & 13\% & 17\% & 16\% & 17\% \\
\hline
Can provide insights but cannot directly assist in decision-making & 13\% & 9\% & 17\% & 22\% & 19\% & 26\% & 22\% & 19\% & 26\% & 32\% & 29\% & 35\% \\
\hline
AI can handle independently without human intervention & 2\% & 0\% & 4\% & 0\% & 0\% & 0\% & 0\% & 0\% & 0\% & 0\% & 0\% & 0\% \\
\hline
AI does not contribute to the process & 0\% & 0\% & 0\% & 4\% & 3\% & 4\% & 4\% & 3\% & 4\% & 0\% & 0\% & 0\% \\
\hline
\end{tabular}
\end{table*}

\subsection{\textbf{RQ3: Challenges and Opportunities of using AI for RE}}
The key challenges and opportunities of using AI for RE are described below. Instead of presenting direct quotes, explanations based on the open-ended survey responses are given, without participant identifiers, since individual attribution was not required for the analysis. The thematic analysis codebook containing the quotes and corresponding themes can be found in \href{https://doi.org/10.5281/zenodo.15648803}{Zenodo}~\cite{b23}.
\vspace{5pt}

\noindent\textbf{Challenges in using AI for RE:} 

\noindent\textit{\textbf{Knowledge and cognitive limitations:}} Participants (15/55) reported that AI models lack deep domain understanding and contextual understanding that is essential for RE tasks. Participants claimed that, in requirements elicitation, \textit{"AI models often lack deep industry expertise, domain-specific knowledge, regulatory knowledge, and organizational memory crucial for requirements elicitation"} and they \textit{"struggle to understand unstated or implicit requirements and tend to rely on existing cases, making them unsuitable for novel or groundbreaking products where clients themselves are uncertain about their needs."} Similarly, in analysis, AI models struggle with interpreting vague and complex requirements, lack deep domain knowledge, and produce unreliable insights from the requirements elicited. AI may miss conflicts and struggle with changing requirements. Current AI models have limited interpretation capabilities (considering multiple angles for the given requirement) for complex requirements analysis. Without adequate product context, AI tends to generate generic solutions rather than product-specific requirements. This indicates that AI's reliance on pattern matching from training data creates significant gaps when dealing with tacit organizational knowledge and novel problem domains that require creative problem-solving beyond existing use-cases.

\noindent\textit{\textbf{Communication and human interaction challenges:}}
Participants (6/55) stated that AI cannot replicate the interpersonal skills and relationship-building capabilities essential for stakeholder engagement, which is a key requirement in RE tasks. In elicitation, AI cannot build rapport, establish trust, interpret non-verbal cues, or navigate politically sensitive requirements, which are essential for effective stakeholder engagement. Participants claimed that, \textit{"while AI can follow scripts, it cannot dynamically adapt questioning based on emerging insights, support creative techniques such as brainstorming, or pivot when new context emerges."}  In specification, explaining requirements to AI systems requires more time investment to achieve a better understanding of the specification. In validation, human critical thinking and judgment are highly required which AI cannot replace. This indicates that RE remains fundamentally a social process requiring emotional intelligence, adaptability, and relationship management skills that current AI cannot provide.

\noindent\textit{\textbf{Quality and accuracy concerns:}}
Participants (8/55) reported issues with AI creating inaccurate, incomplete, or generic outputs that compromise project quality. In elicitation, AI-generated requirements often have inaccuracies and incompleteness, compromising project outcomes. Without proper prompting techniques, AI models could generate non-reproducible results, causing product instability. In analysis, sometimes, when the requirement is refined by an AI model, it will not match the real requirement (stakeholder needs and intentions). Participants emphasized that, \textit{"AI cannot effectively weigh political considerations, business implications, or resource constraints, leading to suggesting technically optimal solutions that are politically unfeasible or misaligned with the organizational goals."} In specification, AI models tend to transform specific requirements into generic specifications due to their reliance on generalized training data. AI is non-deterministic by nature, and the format of output is not structured by default; hence, finding a proper prompting technique to generate structured specification documents is a challenge. This indicates that AI's probabilistic nature and training on generalized datasets can often produce contextually incorrect solutions that require human oversight.

\noindent\textit{\textbf{Technical and implementation barriers:}}
Participants (10/55) highlighted different technical and integration challenges that hinder effective AI adoption in SE tasks. In elicitation, current AI models struggle to provide reliable solutions when dealing with complex business logic or highly intricate decision-making due to functional limitations. Organizations lack adequate training on effective AI tool usage for elicitation. Participants claimed that, \textit{"integration with existing processes, managing bias in AI tools, scalability, user trust, ethical considerations, and cost also pose difficulties."} AI also struggles to recognize connected information across different inputs, requiring multiple regenerations to achieve a comprehensive understanding. The token limit in many LLM models affects information processing, requiring effective management of this constraint for accurate documentation. Technical constraints and workflow integration challenges suggest that current AI tools require significant organizational adaptation to be effective enough to support RE tasks.

\noindent\textit{\textbf{Data and Methodological challenges:}}
Participants (4/55) highlight the training data limitations specific to SE requirements that could limit the capability of AI models in RE tasks.
Participants emphasized that, in elicitation, \textit{"incomplete training data limits AI effectiveness. AI-generated requirements rely on common data sets, causing omission of core or unique requirements."} Improper use of AI tools, such as relying on LLMs instead of direct customer engagement, undermines requirement quality.AI struggles to provide industry-specific requirements (e.g., healthcare vs finance) without extensive customization. In specification, AI models may find it difficult to maintain clear links between raw requirements and final documentation. AI cannot currently tailor specifications to specific client requirements, particularly when constrained by budget or clarity issues. This indicates the importance of domain-specific training data along with general domain data while ensuring appropriate tool deployment strategies.

\noindent\textit{\textbf{Governance, security and compliance challenges:}}
Participants (7/55) expressed concerns about risk management, data protection, and regulatory compliance when using AI tools for RE tasks. In elicitation, using AI models can cause privacy, security, and legal risks for organizations. AI struggles with explainability in its decision-making for specific elicitation techniques and has difficulty handling evolving requirements. In specification, using AI raises concerns about privacy issues and information leakage. In validation, validation against regulatory compliance and changing requirements is challenging for AI models to attain with their current capabilities. Participants claimed that,\textit{"while AI can cross-reference requirements against documented regulations, it often misses the subtle compliance interpretations and industry practices that come from years of regulatory interaction, particularly data privacy and consumer protection nuances."} Confidentiality of the requirements can make the validation process challenging. This indicates that organizational policies and regulatory requirements may significantly constrain AI adoption, mainly in highly regulated industries where compliance interpretation requires human judgment.

\vspace{5pt}
While challenges are evident across all RE phases, practitioners also identified clear opportunities for AI integration.

\noindent\textbf{Opportunities in using AI for RE:}

\noindent\textit{\textbf{Efficiency and Automation:}} 
Participants (5/32) highlighted AI's potential to automate routine RE tasks. AI models can generate initial drafts of requirements, summarize stakeholder inputs, and identify potential gaps, thereby streamlining workflows and boosting productivity. AI enables the automation of various elicitation processes. AI models can transform months-long requirement processes into weeks by providing clear technical specifications and phased implementation plans that could reduce miscommunication and delays. Participants claimed that, \textit{"AI can enhance rather than replace stakeholder validation meetings by pre-identifying issues, allowing meeting time to focus on confirming findings and decision-making, improving both efficiency and stakeholder engagement."} This indicates that practitioners view AI primarily as a productivity multiplier that can speed up routine RE processes while maintaining human oversight and decision-making authority.

\noindent\textit{\textbf{Intelligent analysis and pattern recognition:}} 
Participants (5/32) agreed on AI's ability to process a large amount of data and identify patterns that would be difficult for humans to do manually. AI can analyze past project data and user feedback to identify patterns, suggest missing requirements, and detect inconsistencies. Participants emphasized that, \textit{"AI tools can use NLP to analyze stakeholder interviews, identify key requirements, flag ambiguities, cross-reference documentation, and detect gaps while maintaining human expert oversight for accuracy."} Various trained models can be used to analyze usage statistics and user behavior from data sources such as A/B testing data that will be difficult to analyze manually. These data can be a rich source to match user behavior with organizational priorities while performing requirements analysis. AI can streamline documentation by processing diverse inputs, auto-generating use cases, and ensuring completeness, resulting in faster, accurate, and traceable documentation, thereby reducing human effort. This indicates that practitioners recognize AI's capability for synthesizing insights from diverse data sources while maintaining consistency across complex requirements.

\noindent\textit{\textbf{Knowledge and expertise augmentation:}} 
Participants (7/32) consider AI models as a powerful tool for knowledge and expertise augmentation rather than as a replacement for human expertise. Participants claimed that, \textit{"LLMs excel at combining requirements in novel ways, kickstarting slow processes, providing critical review of existing requirements, and processing large-scale user feedback to propose new features."} AI models can help understand personas, generate questions for stakeholders, create journey maps, and prepare for client meetings to elicit needs and requirements better. AI assistants can help bring out the "obvious" requirements that only the subject matter experts could know. AI can easily identify requirements that are essential but overlooked by non-subject matter expert engineers. With proper guidance, AI models can conduct extensive product research across a vast range of data sources to gather relevant information. AI models can transform non-technical ideas into clear requirements, ask clarifying questions to fill gaps, and provide structured requirement templates, analyze and understand complex legacy codebases, schemas, and documentation to understand current capabilities of these systems, identify integration points, and surface hidden dependencies for better technical feasibility assessment. This indicates that practitioners prefer AI as an intellectual multiplier that can augment human expertise rather than replace it, indicating its potential as a cognitive partner

\noindent\textit{\textbf{Quality and compliance assurance:}}
Participants (8/32) highlighted AI's ability to improve quality control and regulatory compliance processes. Participants highlighted that, \textit{"AI models can efficiently process regulatory documentation, identify regulatory changes, cross-reference requirements, and flag conflicts, while continuously monitoring regulatory developments in real-time and alert requirements teams about potential system impacts due to these changes."} AI can help in the clarification of vague terms, detect conflicts in requirements, prioritize key features, check for compliance, and suggest improvements. If AI models are trained adequately, they can provide comprehensive regulatory coverage and technical feasibility analysis at a better scale and speed, while addressing both functionality and risk management needs, mainly for sectors including banking and defense, where both functionality and risk management are important. AI models can ensure whether stakeholder needs are met, identify conflicts, verify regulatory compliance, align stakeholders, and recommend improvements. AI can improve validation by analyzing documents for inconsistencies, comparing them with historical data, and automatically generating test cases. It can also learn from past validations, reducing manual effort and ensuring high-quality requirements. This shows that AI models, e.g., LLMs, have potential usage in systematic quality assurance tasks that involve cross-referencing and consistency checking across large sets of documents. 

\noindent\textit{\textbf{Documentation and Organization:}}
Participants (5/32) highlighted the potential of AI models in improving documentation quality and thus improving organizational efficiency. AI models can generate technical specification documents that can save significant time, requiring only developers to review and edit these documents. Models can help in grouping and classifying similar requirements and in identifying those within the same functional areas. Participants pointed out that, \textit{"AI uses NLP to analyze stakeholder inputs, identify key requirements, detect ambiguities, and translate findings into detailed documentation."} LLMs ensure accuracy, consistency, and efficiency in specifications. AI models can support the automatic categorization of functional and non-functional requirements, streamlining the initial specification process. This indicates that practitioners view AI as effective for structured documentation tasks that benefit from consistency and systematic organization.

\section{Discussion}

Our survey of 55 software practitioners reveals two clear takeaways: first, AI is now widely used across all four RE phases, and second, practitioners favor HAIC over full automation or passive validation. The most significant finding is the emergence of the collaborative human-AI partnership (HAIC) for decision-making across all RE phases, ranging from 49.2\% to 60.5\%, while full AI automation (3.8\%-7.6\% across phases) remains limited. The findings indicate that AI models (e.g., LLMs) have evolved beyond experimental usage into a mainstream tool in the RE process, as stated by ~\cite{b1,b2,b3}. This indicates that rather than a complete automation or human-only approach, HAIC has more potential in RE tasks~\cite{b12,b13}. This opens the possibility for the AI models to evolve as a collaborative partner rather than as a replacement or simple validation tool, as stated by Hamza et al~\cite{b12}. 

Our findings align with Parasuraman et al.’s levels-of-automation theory~\cite{b11}, which argues that intermediate automation, which uses human oversight while leveraging AI capabilities, often yields the best results. Unlike prior work that focused on LLM benchmarks or technical feasibility~\cite{b1,b2}, we highlight real-world usage patterns of AI in RE and emphasize the barriers that restrict full automation. The emphasis on domain knowledge gaps and challenges in context-specific understanding in AI models indicates the limitation of using general-purpose AI models in specialized domains and specialized tasks such as RE. This implication is in line with Ling et al.~\cite{b11} on the challenges of directly applying LLMs to solve sophisticated problems in specialized domains.

High HAIC rates might reflect current tool limitations-practitioners collaborate simply because existing AI cannot reliably handle end-to-end RE tasks. Future studies should examine whether more mature, domain-specific models shift these balances. Notably, although this study shows a strong correlation between AI perception and usage in RE, 39.1\% of non-users with favorable attitudes towards using AI for RE still refrain from adopting it, suggesting that access, skill gaps, and cost issues, rather than perception, limit AI adoption. 

The limited adoption of fully autonomous AI (3.8\%-7.6\% across phases) despite the positive attitudes toward AI indicates that practitioners have a careful approach to automation in RE. This caution can be mainly due to the high-stakes nature of requirements decisions, regulatory and compliance considerations, and the recognition that RE involves complex stakeholder negotiations that require human judgment. This indirectly emphasizes that human aspects, including domain knowledge and communication skills, are critical in RE phases as stated by Hidellaarachch et al in ~\cite{b14}. Given practitioners' cautious approach towards fully autonomous AI, collaborative AI approaches hold greater promise for adoption. Rather than focusing solely on removing barriers to autonomous AI, organizations could benefit more from developing collaborative AI solutions that address practical implementation issues, including improving access to AI tools designed for HAIC, building skills for effective AI partnership, and managing costs for collaborative AI systems, to mention a few.

This study also highlighted how practitioners approach responsible practices while using AI for RE. Practitioners show AI risk awareness through high reactive oversight (81.2\% human review) but avoid systematic governance practices (37.5\% risk assessment/documentation), which prioritizes immediate productivity over long-term accountability and organizational learning. This governance gap represents both a significant organizational risk (blind spots, stakeholder trust issues, scalability issues) and a major opportunity. Organizations that solve the AI integration challenges through systematic governance approaches will likely achieve more sustainable, scalable, and trustworthy AI adoption. 

The consistency of challenges across RE phases, such as context limitations, domain knowledge gaps, and quality issues, indicates that these are not phase-specific problems but rather a limitation in AI technologies when used in specialized areas such as RE. Given these challenges, the preference for a collaborative model in RE tasks shows a potential in redefining the role of AI as collaborative partners where practitioners can guide the AI models while applying human judgment, tacit knowledge, and domain expertise to ensure an effective RE process.

\noindent\textbf{Implications for Practitioners, Researchers, and AI developers:} \textbf{Practitioners} can benefit from focusing AI integration efforts on analytical tasks where collaboration adds value, while maintaining human oversight for activities involving stakeholder communication. Understanding how AI capabilities can enhance RE means recognizing that human expertise is still crucial for handling complex business logic and decision-making, ensuring accuracy, context awareness, and sound judgment. Organizations may consider implementing AI usage audits regularly, maintaining transparent stakeholder communication about AI involvement, while establishing comprehensive governance frameworks alongside strategic investments for responsible AI incorporation. \textbf{Researchers} are recommended to investigate barriers to AI adoption in RE and develop frameworks for optimizing HAIC while conducting in-depth research to study the effectiveness of AI collaboration in RE-specific phases. \textbf{AI tool developers} may benefit from designing tools that enhance collaboration rather than automation, prioritizing features that augment human capabilities rather than fully automating RE tasks. Using trust-building mechanisms could increase AI adoption while focusing more on technical and ethical capabilities, including advanced natural language processing, context-aware algorithms, structured data cleaning, transparency and explainability, and adherence to ethical guidelines.
\vspace{-0.15cm}
\section{Conclusion}

Our study demonstrates that AI in RE works best not as an autonomous agent, but as a collaborator guided by human expertise, indicating HAIC as an optimal approach for using AI for RE. This study also reveals a concerning gap between AI adoption and its responsible use. This further highlights the urgent need for structured governance frameworks to integrate AI in RE tasks. As AI tools become more capable, embedding them within structured HAIC frameworks and governance models will be essential to unlock their full potential without sacrificing human judgment at the heart of RE. In conclusion, the success of the HAIC approach in RE depends on addressing practical barriers beyond perception, developing appropriate governance frameworks, and creating domain-specific AI tools that understand RE contexts. For future research,  we recommend longitudinal case studies to observe how HAIC evolves as tools improve, and controlled experiments to measure the impact of specific governance practices on RE outcomes.

\end{document}